\documentclass[letterpaper, 10 pt, conference]{ieeetran}  
\IEEEoverridecommandlockouts                  
\usepackage{graphicx}  
\usepackage{subfigure}
\usepackage{multirow}
\usepackage{textcomp}
\usepackage{amsmath}
\usepackage{ifthen}
\usepackage{amssymb}
\usepackage{amsfonts}
\usepackage{subcaption}
\usepackage{colortbl}
\usepackage{algorithm,comment}
\usepackage{algpseudocode}
\usepackage{hyperref}
\usepackage{mathtools,color}
\usepackage{xcolor}
\usepackage{array}

\title{\LARGE \bf
Towards Enabling Learning for Time-Varying finite horizon Sequential Decision-Making Problems*
}

\author{Dhananjay Tiwari$^{1}$, Salar Basiri$^{2}$ and Srinivasa Salapaka$^{3}$% <-this % stops a space
\thanks{*This work was supported by National Aeronautics and Space Administration under Grant NASA 80NSSC22M0070.}% <-this % stops a space
\thanks{$^{1}$PhD Student in Mechanical Science and Engineering, University of Illinois Urbana-Champaign,
        {\tt\small dtiwari2@illinois.edu}}%
\thanks{$^{2}$PhD Candidate in Mechanical Science and Engineering, University of Illinois Urbana Champaign,
        {\tt\small sbasiri2@illinois.edu}}%
\thanks{$^{3}$Professor in Mechanical Science and Engineering, University of Illinois Urbana Champaign,
        {\tt\small salapaka@illinois.edu}}%
}

\newcommand{\mc}[1]{\mathcal{#1}} % short for \mathcal{ . }
\newcommand{\Sinf}{\mathcal{S}_i} % short for infinite-horizon MDP state-space
\newcommand{\Ainf}{\mathcal{A}_i} % short for infinite-horizon MDP action-space
\newcommand{\mb}[1]{\mathbb{#1}} % short for \mathbb{ . }
\newcommand{\mf}[1]{\mathbb{#1}} % short for \mathbf{ . }
\newcommand{\curlyb}[1]{\left\{#1\right\}} % short for curly { . } braces 
\newcommand{\rectb}[1]{\left[#1\right]} % short for rectangular [ . ] braces
\newcommand{\roundb}[1]{\left(#1\right)} % short for round ( . ) braces 

\begin{document}

\maketitle
\thispagestyle{empty}
\pagestyle{empty}

%%%%%%%%%%%%%%%%%%%%%%%%%%%%%%%%%%%%%%%%%%%%%%%%%%%%%%%%%%%%%%%%%%%%%%%%%%%%%%%%
\begin{abstract}
Parameterized Sequential Decision Making (Para-SDM) framework models a wide array of network design applications spanning supply-chain, transportation, and sensor networks. These problems entail sequential multi-stage optimization characterized by states, control actions, and cost functions dependent on designable parameters. The challenge is to determine both the sequential decision policy and parameters {\em simultaneously} to minimize cumulative stagewise costs. Many Para-SDM problems are NP-hard and often necessitate time-varying policies. Existing algorithms tackling finite-horizon time-varying Para-SDM problems struggle with scalability when faced with a large number of states. Conversely, the sole algorithm addressing infinite-horizon Para-SDM assumes time (stage)-invariance, yielding stationary policies. However, this approach proves scalable for time-invariant problems by leveraging deep neural networks to learn optimal stage-invariant state-action value functions, enabling handling of large-scale scenarios. This article proposes a novel approach that reinterprets finite-horizon, time-varying Para-SDM problems as equivalent time-invariant problems through topography lifting. Our method achieves nearly identical results to the time-varying solution while exhibiting improved performance times in various simulations, notably in the small cell network problem. This fresh perspective on Para-SDM problems expands the scope of addressable issues and holds promise for future scalability through the integration of learning methods.
\end{abstract}

\section{Introduction}
\subsection{Motivation}
Parameterized Sequential Decision Making (para-SDM) problems  cover a vast range of spatial network logistics and planning application domains such as last-mile delivery \cite{last_mile},
 supply chain networks \cite{POURNADER2021108250},
 power grids resource allocation \cite{power}, internet of things \cite{iot}, Unmanned Aerial Vehicle (UAV) trajectory optimization \cite{joint_uav,ourAIAA}, industrial robot-resource allocation, and small cell network design in 5G networks \cite{5g}.   These problems include large subclasses of problems such as Markov Decision Processes (MDPs), reinforcement learning (RL), clustering, resource allocation, scheduling, and routing problems, and  data aggregation, classification, and clustering algorithms.

These problems have similar descriptions as dynamic programs or markov decision problems (MDPs) in the sense that they are  characterized by states, representing the system's configurations, and stages, representing decision (control)  points over time. The dynamics describe how the system evolves from one state to another, while the sequential cost function assigns a cost to each action taken at each stage. The main difference from dynamic programs or MDPs is that the states, control, and cost functions in para-SDMs depend on parameters, which themselves are decision variables. Consequently para-SDMs require {\em simultaneously} determining  {\em policies} (characterized by sequence of controls or actions)  and {parameters} (characterized typically by real variables), while incorporating application-specific capacity and exclusion constraints, and while respecting the dynamical evolution of the network. This difference is critical since the underlying optimization problems need not be, and typically are, not MDPs. In fact many of para-SDM prblems are NP-hard.  In this regard, they generalize well studied Markov Decision Process (MDP) problems, where only optimal policies are sought. 

\begin{figure}[tbhp!]
    \centering
\begin{tabular}{c}
    \includegraphics[width=0.750\columnwidth]{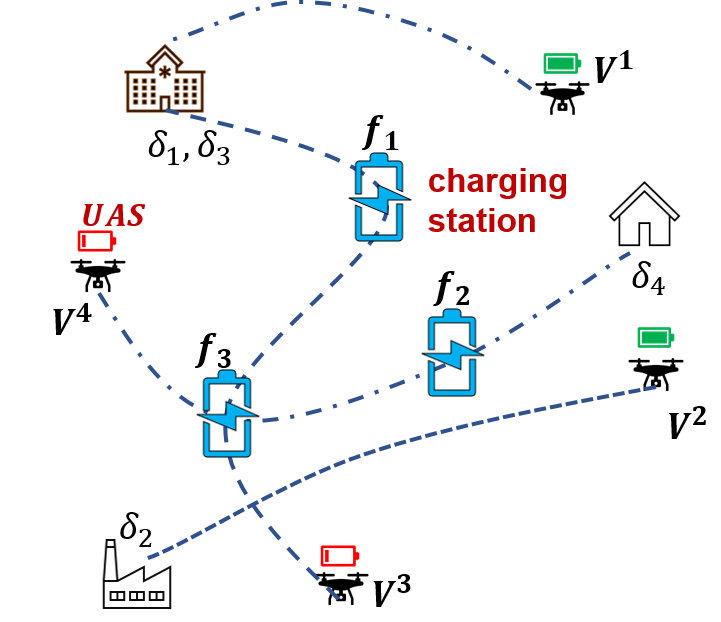}\cr (a)\cr
    \includegraphics[trim={1.5cm 3.5cm 6.5cm 3.5cm},clip,width=0.750\columnwidth]{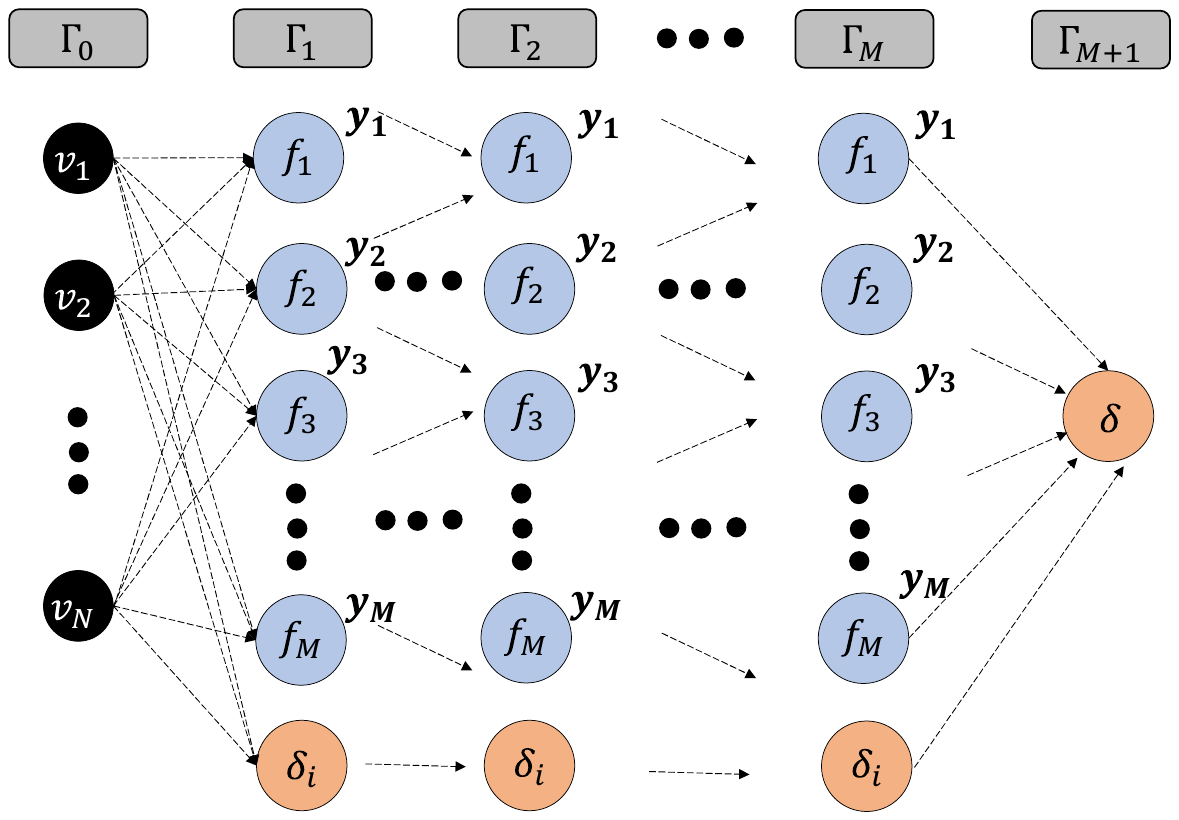}\cr (b)
\end{tabular}
    \caption{Facility-Location Path-Optimization Problem in UAV transport network: (a) This problem shows $N=4$ UAVs $V_i$ each with different amount of initial battery charge values; and their target destinations are $\delta_i$. The objective is to determine the locations $y_j$ of charging UGV facilities $f_j$ as well as the routes (sequence of UGVs) of each UAV  such  that the cumulative travel-distances of all UAVs to their respective destinations is minimized. Note that UAV's have to hop onto UGVs for charging before if they have insufficient charge to reach their destination. (b) A stage-wise depiction of the FLPO problem. Here the FLPO path is reinterpreted in this graphical network, Here the set of possible  UAV trajectories is divided into stages, where states in each stage comprise the facilities and the destination. A path comprises a sequence of states that a particular UAV takes. The objective is to determine the shortest path, while also determining the parameter associated (location of the facility) with each state.}
    \label{fig:FLPO_UAV}
\end{figure}
The main features of the para-SDM are easily explained in terms of a facility-location path-optimization (FLPO) problem that we addressed in \cite{ourAIAA}. In this problem, we have a set of unmanned air vehicles (UAVs) (see Figure \ref{fig:FLPO_UAV}) that enter a domain each with a certain battery charge and a target destination to go to. The network also comprises unmanned ground vehicles (UGVs) as charging stations, where the UAVs can get charged if necessary. The underlying problem is to determine the sequence of UGVs each UAV needs to go to so that every UAV reaches its destination with shortest distance enroute. This problem can be cast as a sequential-decision making problem as shown in Figure \ref{fig:FLPO_UAV}-b, where the goal is to determine the {\em shortest path} for each UAV while ensuring that each UAV is not ever without charge during its travel. This path-optimization problem by itself can be cast as a dynamic program and its stochastic variant as an MDP. However in the problem that we address, we  have to simultaneously determine the {\em shortest path} and the {\em UGV (facility) locations} so that the total travel distance is minimized. These simultaneous facility-location and path-optimization (para-SDM) problems are  much harder problems (NP-hard) and in fact cannot be cast as pure MDP problems. MDPs are special cases of para-SDM problems, when the parameters (charging station locations) are fixed.  Para-SDM problems can have added complexities in terms of topological, capacity, communication, and dynamic (possibly stochastic) constraints on states, actions, and parameters.

There is scant research that address para-SDM problems. Even though extensive literature exists on the facility location problem (FLP) and the shortest-path problem (MDP) individually \cite{path_plannin_review,flp_review}, there is limited research that addresses 
 simultaneous FLPO (para-SDM) problems.  Most existing solutions address these problems sequentially, where facility locations are decided first and then policies are determined. In our previous works, we have addressed two subclasses of para-SDMs - A finite-Horizon {\em time-varying} FLPO problem \cite{srivastava2020simultaneous} and infinite horizon {\em time-invariant} para-SDM problems \cite{srivastava2021parameterized}. Each of these subclasses represent a large class of application areas. The solution procedures proposed, though having, a lot of commonalities have some basic conceptual differences.

\subsection{Background}

 In \cite{srivastava2020simultaneous}, the underlying para-SDM (finite-horizon time-varying FLPO) problem is primarily viewed as a variant of NP-hard facility location problem. A FLP essentially entails determining locations $\{y_j\}_{j=1}^M$ of $M$ \textit{facilities} in a domain of $N\gg M$ \textit{users}, whose locations $\{x\}_{i=1}^N$ are known. The objective is to determine the facility locations $\{y_j\}_{j=1}^M$ such that the cumulative distance of users from their nearest facilities is minimized; that is $\sum_i\sum_j \nu_{j|i} d(x_i,y_j)$ is minimized. Here $\nu_{j|i}\in\{0,1\}$ is a binary variable where $\nu_{j|i}=1$ only when $j$th facility is closest to the $i$th user.    In \cite{srivastava2020simultaneous}, a FLPO problem is posed as a FLP problem as 
 \[\min_{\{y_j\},\{\nu_{j|i}\}}\sum_i\sum_j \nu_{j|i} d(x_i,\gamma_j),\] 
where $\gamma_j$ depicts a path  - a sequence of facility locations; here $\gamma_j(k)=y_\ell$ implies that $k$th facility in the route taken by the user is at location $y_\ell$. Accordingly the solution proposed in \cite{srivastava2020simultaneous} extends a maximum-entropy-principle (MEP) based solution \cite{rose1998deterministic} used to solve FLP. Here the main concept is to replace binary variables $\nu_{j|i}$ by probability distributions  $p_{j|i}=\prod_k p_k(y_j|y_k)$, and determining these distributions by using MEP (described in Section II). One of the biggest advantages of using this framework is that the distributions $p_k(y_j|y_k)$ can be explicitly solved for in terms of facility locations $\{y_j\}$.  These solutions are time-varying in the sense that the probability  $p_k(y_s|y_r)$ of choosing $s$th facility in the $(K+1)$th stage of the route given that the user location is at $r$th facility at $k$th stage depends on the stage $k$. Since these distributions  are obtained explicitly, there is a substantial reduction in the optimization effort, since now the decision variables are effectively only facility locations $\{y_j\}$ instead of facility location and route variables. This reduction of variables enable FLPO solutions for reasonably large networks; However for very large networks of nodes and facilities, optimizing the free energy function with respect to facility locations may not be computationally efficient, or even feasible.

% To manage the complexity of these problems, dimensionality reduction techniques are often employed, with Convolutional Neural Networks (CNNs) being a commonly used method within the learning methodology (\textit{any references here?}).
In our previous work \cite{srivastava2021parameterized}, we view the underlying infinite-horizon time-invariant para-SDM problem as a variant of a MDP problem. Here we consider a MDP formulation, where shortest paths are sought from each initial state such that the cumulative cost across the paths are minimized. The main difference here is that the states and the cost functions are characterized by parameters $\{y_j\}$. Here for any fixed values of the parameters, the underlying path-optimization problem is an MDP problem.  Accordingly here, the parameter values are initialized and fixed, and the best time-invariant path (stationary policy) is obtained by deriving and solving corresponding Bellman equation. Here the policy is stationary in the sense that the prescribed control action when at a state does not depend on the stage $k$. Then the best parameters are obtained by minimizing the value function under this policy. This induces an iterative process, where optimal policies and parameters are found by successively fixing them one after the other.  This process by itself does not scale easily to problems with large number of states. In fact, the calcuations here are numerically expensive since they require {\em both} policy as well as parameter computations.  However data-driven learning based tools have proved very successful in making these solutions scalable. Here  machine learning schemes are employed to learn either the value function derived by the Bellman dynamic programming (DP) equation or the policy directly \cite{bertsekas1996neuro, littman1996algorithms}. One approach involves utilizing function approximation techniques, such as deep neural networks (DNNs), to approximate the value function or policy. By leveraging the representational power of neural networks, researchers can effectively handle the high-dimensional state and action spaces. However, these methods are tailored for infinite horizon, time-invariant Bellman equations that yield stationary policies, rendering them less effective for time-varying problems such as FLPO problems. 
% (\textbf{So Amber's ParaSDM paper uses Q-learning to learn state-action value function and gradients via G-iteration. Is Q-learning equivalent to Neural networks? There's no explicit mention of CNNs. As far as I understand, Q-learning is an iterative process to approximate these functions}).{\color{blue} is it okay now?}\textbf{I don't think DNN is used anywhere - Amber's paper is tabular approximation using Q-learning. Mustafa's thesis has DNNs for large state and action spaces but it is not published anywhere}

\subsection{Our Contribution}
In this article, we present a methodology for reinterpreting finite-horizon, time-varying FLPO problems as equivalent stationary, infinite-horizon SDMs via a topographical transformation, in which the states, actions and transition costs are dependent on a set of parameters which are deemed as secondary optimization variables. The main concept is to extend (lift) the state-space in the finite-horizon time-varying FLPO problem and convert it into a time-invariant problem.
% More precisely a $K$-stage FLPO problem with $M$ states in each stage is converted into a time-invariant infinite-horizon problem, where each stage has $KM$ states.
The two-fold optimization problem becomes determining the optimal state-action parameters as well as the optimal stationary policy. The smaller dimensioned time-varying policy for the original FLPO problem, which is quite challenging to obtain in general, can then be inferred from the stationary policy of the infinite-horizon problem. This non-trivial transformation enables learning in the presence of large networks of spatial nodes and facilities which is crucial for numerous real-world applications, resulting in solutions that can effectively scale with the number of states and actions. 
The transformation offers more than just scalability benefits. For instance, in scenarios like UAV transportation, where transition costs between nodes and facilities are often unknown or transitions between spatial coordinates are probabilistic (e.g., due to emergencies or accidents), modeling FLPO problems as MDPs enables us to incorporate these crucial and practical uncertainties. Moreover, it provides us with incredible flexibility to address a broader range of problems in which the unknown variables exhibit time-dependent behavior, as seen in UAV routing and scheduling problems. In this work, we demonstrate via simulations that the numerical solutions obtained from the finite-horizon FLPO problems using techniques from \cite{srivastava2020simultaneous} and the reinterpretted infinite-horizon solutions are very close. 

\section{MATHEMATICAL OVERVIEW}

In this section, we provide a summary of the time-varying FLPO problem \cite{srivastava2020simultaneous} and the time-invariant Parametrized SDM \cite{srivastava2021parameterized}, briefing their solution methods rooted in the Maximum Entropy Principle \cite{rose1998deterministic}. 
A comprehensive understanding of these formulations helps understand the scalability issues with the original FLPO framework and how transitioning to an infinite horizon framework enables it through machine learning techniques.
% Our aim in this paper is to reinterpret the FLPO, typically defined within a finite time frame, as a Parametrized SDM extending into an infinite horizon. \textit{Emphasize: Understanding these formulations is important to understand our work}

\subsection{Facility Location and Path Optimization Problems}
\label{subsec:FLPO}

% In this section, we provide a brief overview of the FLPO problem modeled within a stage-wise framework and the solution approach for non-convex objective function \cite{srivastava2020simultaneous}. 
% by overcoming the initialization biases {\color{red}ambiguous...try clarifying}\cite{srivastava2020simultaneous}. 
% The solution approach is based the Maximum Entropy Principle (MEP) for which readers are encouraged to refer to the articles.

\begin{figure}
    \centering
    \includegraphics[scale=0.27]{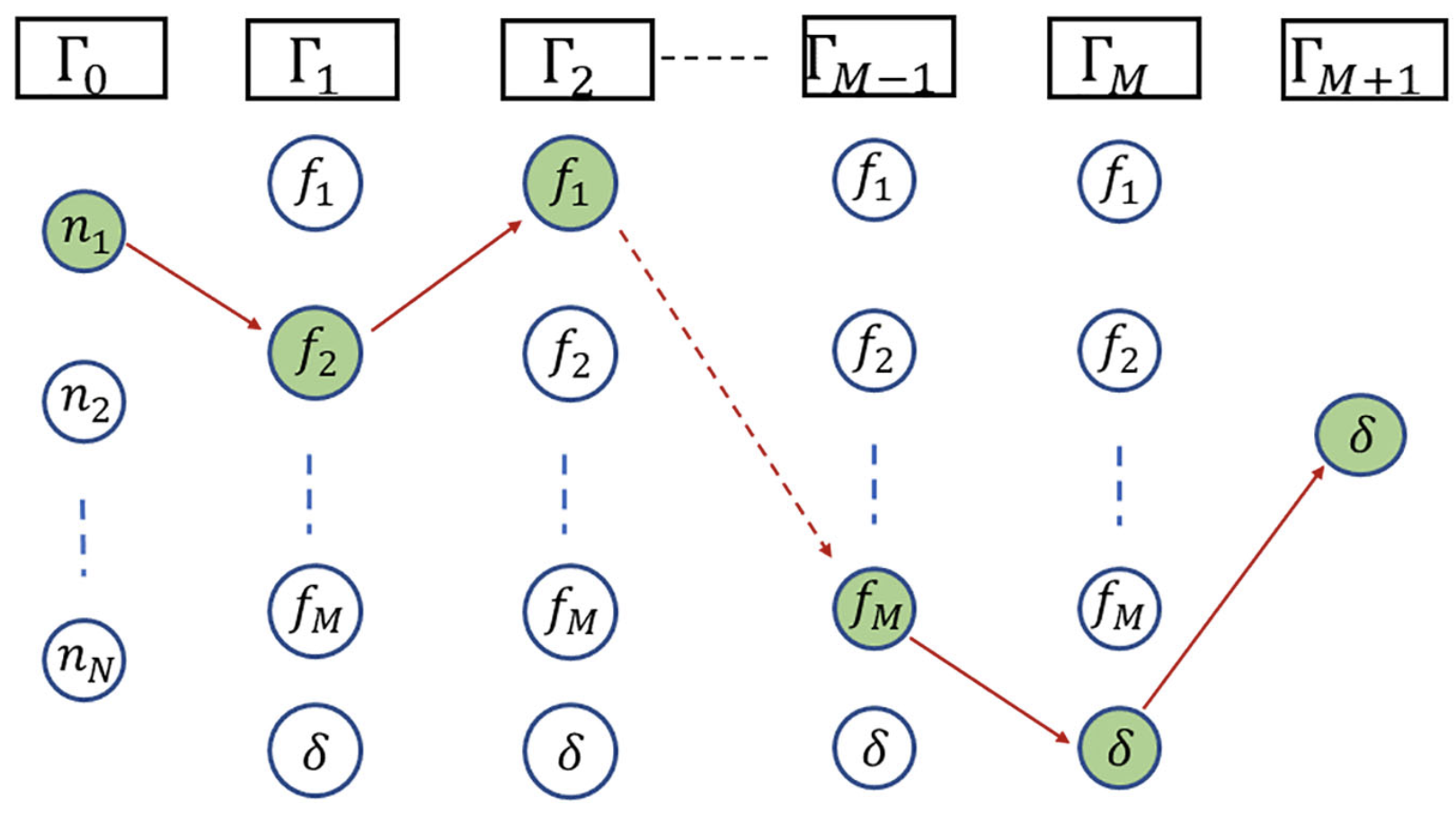}
    \caption{Stagewise FLPO Architecture. Green nodes demonstrate a transportation path from node $n_1$ to $\delta$ via stages $\curlyb{\Gamma_k}_{k=1}^{M+1}$ }
    \label{fig:FLPO_stagewise}
\end{figure}

In its simplest form, the FLPO problem is characterized by overlaying a network of $M$ facilities $\left\{f_j\right\}_{j=1}^M$ at locations $\curlyb{y_j}_{j=1}^M$, over a large number of nodes $\left\{n_i\right\}_{i=1}^N$, fixed at locations $\curlyb{x_i}_{i=1}^N$, and finding a path from each node to a destination center $\delta$ through the facilities $f_j$ such the total cost of transportation along the routes is minimized. 
% The nodes $n_i$ are fixed at the locations $x_i \in \mb{R}^q$, $z \in \mb R^q$ is the destination location and the facilities $f_j$ have unknown parameters $y_j \in \mb{R}^q$. 
In \cite{srivastava2020simultaneous}, a stagewise architecture is proposed to model the problem as an SDM in finite horizon (see Figure \ref{fig:FLPO_stagewise}). 
Stage $\Gamma_0$ consists of all the nodes $\curlyb{n_i}$, stages $\curlyb{\Gamma_k}_{k=1}^M$ consist of all the facilities $\curlyb{f_j}$ and the destination center $\delta$ and $\Gamma_{M+1}$ is the termination state.
A transportation path from nodes to destination through facilities is denoted by $\gamma = \roundb{\gamma_1, \gamma_2, \dots, \gamma_M}, \gamma_k \in \Gamma_k, \forall k$
% \curlyb{\roundb{\gamma_1, \gamma_2, \dots, \gamma_M}: \gamma_k \in \Gamma_k, \forall k = 1,2,\dots,M}$ represent the set of all transportation paths to the destination center $\delta$ from $\gamma_0 \in \Gamma_0$, 
and $d(\gamma|\gamma_0) = \sum_{k=0}^M d_k(\gamma_k, \gamma_{k+1}), \gamma_k \in \Gamma_k \forall k$ is the cost along $\gamma$ starting at $\gamma_0 \in \Gamma_0$,
where the metric $d_k(\gamma_k, \gamma_{k+1})$ is stagewise cost dependent on the network parameters. The objective is to minimize the cumulative cost $D = \sum_{\gamma_0 \in \Gamma_0} \rho_{\gamma_0}\sum_{\gamma \in \mc G} \nu(\gamma|\gamma_0) d(\gamma|\gamma_0)$ with respect to $\curlyb{y_j}$, where $\nu(\gamma|\gamma_0) = 1$, if $\gamma = \textrm{argmin}_{\gamma}d(\gamma|\gamma_0)$, $\nu(\gamma|\gamma_0) = 0$ otherwise,
% \begin{align}
%     \min_{\curlyb{y_j}} D & := \sum_{\gamma_0 \in \Gamma_0} \rho_{\gamma_0}\sum_{\gamma \in \mc G} \nu(\gamma|\gamma_0) d(\gamma|\gamma_0) \label{eq:distortion_binary}\\
%     \textrm{where } \nu(\gamma|\gamma_0) & = \begin{cases}
%         & 1, \textrm{ if } \gamma = \textrm{argmin}_{\gamma\in \mc G} d(\gamma|\gamma_0) \\
%         & 0, \textrm{otherwise}
%     \end{cases}\nonumber
% \end{align}
and $\rho_{\gamma_0}$ is the weight given to node $\gamma_0 \in \Gamma_0$, often implying its strategic importance.
% As mentioned earlier, the optimization problem (\ref{eq:distortion_binary}) has twofold objectives a) finding optimal value of parameters $y_j \in \mb R^q$ b) finding the optimal routes from each node $\gamma_0$ to $\delta$ through $f_j$. 
% Since the choice of an optimal route $\gamma$ from $\gamma_0$ is defined by $\nu(\gamma|\gamma_0) \in \curlyb{0,1}$, which depends on the facility locations, these objectives are coupled and lead to a $\mc N \mc P - $hard mixed integer programming problem. 
The above optimization problem is an NP-hard mixed integer programming problem with binary decision variables $\nu$ and continuous decision variables $\curlyb{y_j}$.
% Solving the problem sequentially first with already existing FLP algorithms and then path finding algorithms yields suboptimal solutions due to presence of mulitple poor local minima. 
However, the method proposed in \cite{srivastava2020simultaneous} relaxes the objective by replacing the binary variables with a probability distribution $p(\gamma|\gamma_0)$ over all the paths $\gamma$, and minimizes the modified objective $$F := D - \frac{1}{\beta} H$$ where the regularizing term  
% $D = \sum_{{\gamma_0}} \rho_{\gamma_0} \sum_{\gamma \in \mc G} p(\gamma|\gamma_0) d(\gamma|\gamma_0)$ 
% and the Shannon's entropy of the distribution $p(\cdot|\gamma_0)$, 
$H := -\sum_{\gamma\in\mc G} p(\gamma|\gamma_0) \log p(\gamma|\gamma_0)$ is the Shannon's entropy of the distribution $p(\cdot|\gamma_0)$.
% \cite{rose1998deterministic, rose1991deterministic}. 
% The modified objective $D$ becomes
% \begin{align*}
%     D = \sum_{{\gamma_0}} \rho_{\gamma_0} \sum_{\gamma \in \mc G} p(\gamma|\gamma_0) d(\gamma|\gamma_0) 
% \end{align*} 
% resulting into the following unconstrained minimization of
% $F := (D-D_0) - \frac{1}{\beta} H$ with respect to the associations $p(\gamma|\gamma_0)$ and the parameters $\curlyb{y_j}$
% at varying values of the Lagrangian parameter $\beta$ from zero to infinity. 
This regularization is inspired from \textit{Deterministic Annealing} (DA) in clustering problems \cite{rose1998deterministic, rose1991deterministic}. Applying the \textit{law of optimality} we rewrite the probability association $p(\gamma|\gamma_0)$ into stagewise associations $p_k(\gamma_{k+1}|\gamma_k)$ as, $p(\gamma|\gamma_0) = \prod_{k=0}^{k=M}p_k(\gamma_{k+1}|\gamma_k)$ and determine them by setting $\rectb{\partial F/\partial p_k(\gamma_{k+1}|\gamma_k)} = 0$. This results in the following Gibbs distribution
\begin{align}
    p_k = e^{-\beta d_k} \frac{\sum_{\substack{(\sigma_{k+2}, \dots, \sigma_M): \\ \sigma_{k+1} = \gamma_{k+1}}} e^{-\beta\sum_{t = k+1}^M d_t(\sigma_t, \sigma_{t+1})}}{\sum_{\substack{(\sigma_{k+1}, \dots, \sigma_M): \\ \sigma_{k} = \gamma_{k}}} e^{-\beta\sum_{t = k}^M d_t(\sigma_t, \sigma_{t+1})}}
    \label{eq:gibbs_distribution}
\end{align}
where $p_k = p_k(\gamma_{k+1}|\gamma_k)$ and $d_k = d_k(\gamma_{k+1}|\gamma_k)$. The optimal values of $\curlyb{y_j}$ can be obtained by resubstituting the associations $p_k$ into $F$ and solving for $\curlyb{y_j}$ using standard numerical methods \cite{srivastava2020simultaneous, ourAIAA}.
% A closed form expression \ref has been provided in \cite{srivastava2020simultaneous} for the stagewise cost $d_k$ computed using square Euclidean distance and in \cite{ourAIAA} the optimization is performed using \textit{Powell's method} for $d_k$ augmented with additional penalty terms.
% We call $1/\beta$ as the \textit{temperature} parameter and $F$ is called the \textit{free energy} due to their close analogies to the statistical physics.
This process is performed in an iterative manner where $\beta$ is increased from $0$ to infinity. The solutions of previous iterate are used as initial guesses for the next iterate.
Note that as $\beta\rightarrow\infty$, $F = D - \frac{1}{\beta}H \rightarrow D$, the original cost function. Moreover, from (\ref{eq:gibbs_distribution}) we note that the gibbs distribution $p_k$ goes to binary values $\curlyb{0,1}$ thus recovering the hard associations required for the original FLPO problem.

\subsection{Time-Invariant Parametrized SDM}
\label{subsec:parasdm_timeinvariant}
Consider an infinite horizon parametrized sequential decision making problem described by an MDP $\left\langle \mc S, \mc A, c, p, \gamma \right\rangle$, where $\mc S$ is the state-space, $\mc A$ is the action-space, $c:\mc S \times \mc A \times \mc S \rightarrow \mb R$ is the transition cost, $p:\mc S \times \mc A \times \mc S \rightarrow \rectb{0,1}$ is the state-transition probability and $\gamma \in \left(0,1\right]$ is a discounting factor. The states and actions have associated unknown parameters $\curlyb{\xi_s}$ and $\curlyb{\lambda_a}$ on which the transition cost is also dependent. For a given initial state $x_0 = s \in \mc S$, the MDP induces a stochastic process of the form $\curlyb{u_0, x_1, u_2, \dots,}, u_t, x_t \in \mc A, \mc S, \forall t \in \mb Z_{\geq 0}$. The objective is to simultaneously determine 1) a stationary policy $\mu(a|s) \in \curlyb{0,1},$ such that $\mu(a|s) = 1$ if action $a\in\mc A$ is taken at $s\in\mc S$, otherwise $0$ and 2) the optimal value of state and action parameters $\curlyb{\xi_s}$ and $\curlyb{\lambda_a}$ such that the state value function
\begin{align}
    J^{\mu}_{\xi\lambda}(s) = \mb E_{p_{\mu}}\rectb{\sum_{t=0}^\infty \gamma^t c\roundb{x_t(\xi), u_t(\lambda), x_{t+1}(\xi)}|x_0 = s} \label{eq:value_fcn_def}
\end{align}
is minimized $\forall s \in \mc S$. Here $x_t(\xi)$ denotes the state $x_t \in \mc S$ with the associated parameter $\xi_{x_t}$ and $u_t(\lambda)$ denotes the action $u_t \in \mc A$ with the associated action parameter $\lambda_{u_t}$. 
The expectation is with respect to the probability distribution $p_{\mu}(\cdot|s):\omega \rightarrow \rectb{0, 1}$ on the space of all possible paths $\Omega := \curlyb{\roundb{u_t, x_{t+1}}_{t\in \mb Z_{\geq 0}}: u_t\in \mc A, x_t\in \mc S}$. Finding a deterministic policy is a NP-hard problem. However, authors in \cite{srivastava2021parameterized} randomize the policies $\mu$ and use a DA based approach to optimize the free energy $V_{\beta, \xi\lambda}^{\mu}(s) = J^{\mu}_{\xi\lambda}(s) - \frac{1}{\beta}H^{\mu}(s)$, where $H^{\mu}(s)$ is the entropy of $p_{\mu}$ and $\beta$ is the annealing parameter.
The Markov property $p_{\mu}(\omega|x_0) = \prod_{t=0}^\infty \mu(u_t|x_t) p(x_{t+1}|x_t,u_t)$ reduces the problem size and an optimal policy $\mu_\beta^*$ can be obtained by differentiating the state-value function in the following Bellman equation
\begin{align*}
    V_{\beta,\xi\lambda}^{\mu}(s) = \sum_{s,a} \mu_{a|s} p_{ss'}^a \rectb{\bar{c}^a_{ss'} + \frac{1}{\beta} \log \mu_{a|s} + \gamma V_{\beta, \xi\lambda}^{\mu}(s')},
\end{align*}
where $\mu_{a|s} = \mu(a|s), p_{ss'}^a = p(s'|s,a)$ and $\bar c_{ss'}^a = c(s,a,s') + \gamma/\beta \log p_{ss'}^a$. At every $\beta$, the optimal policy $\mu^*_\beta$ satisfies $[\partial V_{\beta,\xi\lambda}^{\mu}(s)]/[\partial \mu_\beta^*(a|s)] = 0$ giving the Gibbs distribution
\begin{align}
    \mu^*_{\beta}(a|s) & = \frac{e^{-\frac{\beta}{\gamma}\Lambda_\beta\roundb{s,a}}}{\sum_{a'}e^{-\frac{\beta}{\gamma}\Lambda_\beta\roundb{s,a'}}}, \label{eq:policy_gibbs}
\end{align}
where $\Lambda_\beta(s,a)$ is the state-action value function which are obtained using the following fixed point iteration
\begin{align}
    \Lambda_\beta(s,a) & = \sum_{s'} p_{ss'}^a \rectb{\bar{c}_{ss'}^a -\frac{\gamma^2}{\beta} \log \sum_{a'} e^{-\frac{\beta}{\gamma}\Lambda_\beta(s,a)}}. \label{eq:stateaction_contraction}
\end{align}
Further the optimal parameters are obtained using the gradient descent scheme $\xi_s^+ = \xi_s^- - \epsilon \sum_{s'}\rho(s')G_{\xi_s}^\beta(s'), \ \eta_a^+ = \eta_a^- - \bar\epsilon \sum_{s'}\rho(s')G_{\eta_a}^\beta(s')$ where the gradients 
$G_{\xi_s}^\beta(s') := \partial V_{\beta, \xi\lambda}^{\mu_i}(s')/\partial \xi_s$  and $G_{\lambda_a}^\beta(s') := \partial V_{\beta, \xi\lambda}^{\mu_i}(s')/\partial \lambda_a$ satisfy the following fixed-point iterations
\begin{align}
    G_{\xi_s}^\beta(s') & = \sum_{a'} \mu_{a'|s'} \sum_{s''} p_{s's''} \rectb{\frac{\partial c_{s's''}^{a'}}{\partial \xi_{s}} + \gamma G_{\xi_s}^\beta(s'')}, \label{eq:gi_stateparams} \\ 
    G_{\lambda_a}^\beta(s') & = \sum_{a'} \mu_{a'|s'} \sum_{s''} p_{s's''} \rectb{\frac{\partial c_{s's''}^{a'}}{\partial \lambda_{a}} + \gamma G_{\lambda_a}^\beta(s'')}.
\end{align}
Similar to the FLPO, DA is performed at increasing values of $\beta$ starting from nearly zero till infinity. At low $\beta$, a stochastic policy corresponding to the uniform distribution over the paths is obtained. The solution from current $\beta$ is used as initial condition for subsequent $\beta$ iterations and as it increases the optimal policy converges to a deterministic policy.\\

\noindent\textbf{Remark}: The computational time of the FLPO problem (discussed in Subsection \ref{subsec:FLPO}) significantly increases as the number of nodes and facilities grows. Similarly, the ParaSDM iterations (as denoted by eq. (\ref{eq:stateaction_contraction}) and eq. (\ref{eq:gi_stateparams})) require more computational time with large state and action spaces. However, these iterations bear similarity to Q-learning iterations found in the literature on reinforcement learning, allowing for approximation of such functions with DNNs.

\section{FLPO AS TIME INVARIANT PARAMETRIZED SDM}
In this section we explore a time-varying SDM model for FLPO, with states, actions, and costs influenced by unknown parameters. We then suggest a time-invariant SDM model achieved through a topographical transformation of the original FLPO framework which results in a stationary time-invariant policy. 
% {\color{red}\textit{Make a big deal out of converting FLPO to paraSDM - critical for scalability. It hasn't been done!}}
% \ref{subsec:parasdm_timeinvariant}.
\subsection{Time-Variant Parametrized SDM Model of FLPO}
% A finite-time horizon MDP of discrete time-horizon length $K$ is formally described by a tuple $\left\langle\mc S, \mc A, c_k, p_k, \rho \right\rangle$ where $\mc S$ is a 
% finite state-space, $\mc A$ is a finite action space, $\forall k = 0,1,\dots,K-1$, $c_k:\mc S \times \mc A \times \mc S \rightarrow \mb R$ is a time-dependent cost function ($c_k(s,a,s')$ denotes the cost of transition from state-action pair $(s,a) \in \mc S \times \mc A$ to the state $s'\in S$), $p_k:\mc S \times \mc A \times \mc S \rightarrow \rectb{0,1} $ is the time-dependent probability distribution ($p_k(s'|s,a)$ denotes the probability of reaching the $s' \in \mc S$ upon taking the action $a \in \mc A$ at the state $s \in S$), $\rho$ is a distribution over the states which specifies how the initial state is generated.

The stage-wise FLPO framework in section \ref{subsec:FLPO} is equivalent to a finite-time horizon SDM of horizon length $K = M$ with state-space
$\mc S = \curlyb{n_1, n_2, \dots, n_N, f_1, f_2, \dots, f_M, \delta}$
and the unknown state parameters $\curlyb{\zeta_s, s\in \mc S}$,
the action space $\mc A = \curlyb{f_1, f_2, \dots, f_M, \delta}$ as the set of all possible directions at any state $s \in \mc S$. 
The initial state is determined by the distribution $\rho: \mc S \rightarrow \rectb{0,1}$  with its support over the set $\curlyb{n_1,n_2,\dots,n_N} \subset \mc S$. 
The state-transition cost $c_k(s,a,s')$, is equivalent to the FLPO stage-transition cost $d_k(\gamma_{k+1} = s'|\gamma_k = s)$ and hence determined by the parameters $\zeta_s$ and $\zeta_{s'}$. 
Define a time-variant control policy $\mu$ as 
\begin{align}
    \mu = \curlyb{\mu_k(a|s) \in \curlyb{0,1} : a \in \mc A, s\in \mc S, k = 0,1,\dots,K-1},
    \label{eq:finite_sequence_of_determinstic_policies}
\end{align}
that determines the action $u_k \in \mc A$ taken at $x_k\in\mc S$ if $\mu_k(u_k|x_k) = 1$, otherwise $0$. 
% Further, for a given state-action pair $(x_k, u_k)$, the state-transition probability $p_k(\cdot|x_k,u_k):\mc S \rightarrow \rectb{0,1}$ determines a state $x_{k+1} \in \mc S$. 
To resemble the FLPO, we consider a deterministic state-action probability transition $p_k$ such that $p_k(x_{k+1}|x_k,u_k) = 1$, if $x_{k+1} = u_k$, otherwise $0$. 
% However, n incorporating the model uncertainties by making $p_k$ stochastic may expand the scope of FLPO. 
At time step $k < K$, and a given state $x_k = s \in \mc S$, the MDP induces a finite path
$    
\omega_k  = \roundb{u_k, x_{k+1}, u_{k+1}, \dots, x_K}, (u_t, x_{t+1}) \in \mc A \times \mc S, \forall t
$.
Let $\Omega_k$ be the set of all the paths of the form $\omega_k, \forall k$.
Note that $\omega_0 \in \Omega_0$ is similar to a path $\gamma \in \mc G$, but with an added flexibility of taking a decision $u_k \in \mc A$ at given $x_k \in \mc S, \forall k$. 
For the policies of the form $\mu$ in (\ref{eq:finite_sequence_of_determinstic_policies}) define the state-value function $J_k^\mu(s), \forall s \in S, \forall k = 0,1,\dots,K-1$
\begin{align*}
    J_k^{\mu}(s) = \mf E_{p^\mu_k}\rectb{\sum_{t=k}^{K-1} c_t(x_t,u_t,x_{t+1}) + c_{K}(x_K,\delta)|x_k = s},   
\end{align*}
where $c_K(x_K,\delta)$ is the terminal cost penalizing the state $x_K$ for terminating away from $\delta$, identical to the final stage cost $d_M(\delta|\gamma_M = x_K)$ in the FLPO. 
The expectation is taken over the distribution $p^\mu_k(\cdot|x_k = s):\Omega_k \rightarrow \rectb{0,1}, \forall k$ which determines the probability of choosing a path $\omega_k \in \Omega_k$ starting at $x_k = s \in \mc S$. 
% Using the Markov property, we can dissociate the probability $p_k^\mu(\omega_k|x_k = s)$ as 
% \begin{align}
    % p_k^\mu(\omega_k|x_k = s) = \prod_{\substack{t=k \\ x_k = s}}^{K-1} \mu_t(a_t|x_t) p_t(x_{t+1}|x_t,u_t) 
    \label{eq:finite_markov_property}
% \end{align}
The FLPO problem is identical to obtaining an optimal policy $\mu^* = \curlyb{\mu_k^*}$ and the optimal parameters $\curlyb{\zeta_s^*}$ such that the cost $\sum_{s} \rho(s)J_0^{\mu}(s)$ is minimized. 
% where value function $V_0^\mu(s), \forall s \in \mc S$ is minimized.
% \begin{align*}
%     \min_{\substack{\curlyb{\mu_k}, \curlyb{\zeta_s}}} \mf E_{p^\mu_0}\rectb{\sum_{t=0}^{K-1} c_t(x_t,u_t,x_{t+1}) + c_{K}(x_K,\delta)|x_0 = s} 
% \end{align*}
As before, a DA based approach can be used by relaxing the problem with stochastic policies of the form
\begin{align*}
    \pi = \curlyb{\pi_k(a|s) \in \rectb{0,1} : a \in \mc A, s\in \mc S, k = 0,1,\dots,K-1},
    \label{eq:finite_sequence_of_stochastic_policies}
\end{align*}
and minimize the cost $\sum_{s} \rho(s) V_{0,\beta}^{\mu}(s)$ using DA, where $V_{k,\beta}^{\mu}(s) = J_k^{\mu}(s) - 1/\beta H_k^{\mu}(s), \forall k$ and $H_k^{\mu}(s)$ is Shannon's entropy of the distribution $p_k^{\mu}$. Since the initial distribution is known, optimization with respect to $\mu$ reduces to optimizing $V_{0,\beta}^{\mu}(s), \forall s$.
% We work with a set of stochastic policies $\Pi$ 
% \begin{align*}
%     \Pi = \big\{\pi & = \big\{\pi_k(\cdot|s):\mc A \rightarrow \roundb{0,1}\big\}, \\ 
%     & s \in \mc S, k = 0,1,\dots,K-1\big\}
% \end{align*}
% and use the Maximum Entropy Principle to determine the distribution $p^\mu_0$ for $\mu \in \Pi$ such that the Shannon's Entropy is maximized and the value function $V_0^\mu(s)$ attains specific values $V_0$
% \begin{align*}
%     \max_{p^\mu_0: \mu \in \Pi} H^\mu_0(s) := & -\mf E_{p^\mu_0}\rectb{\log p^\mu_0(\omega_0|s)}  \\
%     \textrm{s.t.} \ V_0^\mu(s) = & \ V_0
% \end{align*} 
% We apply the markov property in eq. (\ref{eq:finite_markov_property}) and write the resulting Lagrangian, $F_0^\beta(s) = V_0^\mu(s) - \frac{1}{\beta}H_0^\mu(s) = $
% \begin{align*}
%     \mf E_{p^\mu_0} \Bigg[ & \sum_{t=0}^{K-1} c_t(x_t,u_t,x_{t+1}) + \frac{1}{\beta}\log \mu_t(u_t|x_t) \\ & + \frac{1}{\beta}\log p_t(x_{t+1}|x_t, u_t) + c_{K}(x_K,\delta)\left.\right|x_0 = s \Bigg]
% \end{align*}
% When $s$ is drawn according to $\rho$, then $\sum_{s\in S}\rho(s)V_0^\mu(s)$ is identical to the relaxed cost $D = \sum_{{\gamma_0}} \rho_{\gamma_0} \sum_{\gamma \in \mc G} p(\gamma|\gamma_0) d(\gamma|\gamma_0)$.
Such problems can be solved using the well-known finite horizon Bellman programming. The optimal value function $V_k^*, \forall k = 0,1,2,\dots,K-1$ satisfies
\begin{align}
    V_k^*(s) = \min_{a \in \mc A} \sum_{s'\in \mc S} & \mu_k(a|s) p_k(s'|s,a) \cdot \nonumber\\ 
    &\quad \quad \rectb{c_k(s,a,s') + V_{k+1}^*(s')}, \nonumber
\end{align}
% It can be proved that if $\mu_k^*$ attains minimum for each $k$ in the above equation, then the policy $\mu* = \curlyb{\mu_0^*, \mu_1^*, \dots, \mu_{K-1}^*}$ is optimal for the $K-$step problem. 
Unfortunately, in many practical cases an analytical solution of the policies for the finite-horizon DP is not possible or they are numerically expensive to compute. Moreover, the optimization with respect to parameters adds to the computational complexity of the current scenario.
On the other hand, infinite horizon problems represent a reasonable approximation of problems involving large state and action spaces, with comparatively straighforward implementation of stationary optimal policies \cite{bertsekas1996neuro}.
Hence we propose a method to \textit{lift} the above time-varying model of FLPO by augmenting all the stagewise transitions into a single stage transition and visualize the problem as an infinite horizon parametrized SDM as demonstrated in section \ref{subsec:parasdm_timeinvariant}. 
We augment the FLPO stages $\Gamma_0,\Gamma_1,\dots,\Gamma_{M+1}$ into a single stage $\Sigma_t, t \in \mb N$ with facilities belonging to the distinct stages of FLPO treated as distinct facilities in $\Sigma_t$ by adding a superscript corresponding to the stage index. Hence $\Sigma_t$ is represented by 
\begin{align*}
    \Sigma_t = \Gamma_0 \cup \Gamma_1 \cup \dots \cup \Gamma_{M+1},
\end{align*}
and the elements of $\Sigma_t$ are denoted by, $n_i \in \Gamma_0, i =1,2,\dots,N$, $f_j^k \in \Gamma_k, \forall j =1,2,\dots,M, \forall k =1,2,\dots,M$ and $\delta \in \Gamma_{M+1}$. The parameters associated with the nodes are $\curlyb{x_i}_{i=1}^N$, that with facilities are $\curlyb{y^k_j}_{j,k=1}^M$ and $z \in \mb R^q$ is the destination parameter. Note here we have distinguished the unknown parameters for the same facility in an FLPO problem, which provides us additional flexibility for the scenarios where the parameters vary across the stages in an FLPO.
% For instance, scheduling and routing problems in UAV networks can be dealt with by optimizing for multiple time schedule parameters assigned to the same facility.
In order to respect the transitions of the form $\Gamma_k \rightarrow \Gamma_{k+1}$ in the FLPO, we impose topological constraints on the state and action spaces, allowing us to capture all the sequential transitions of the FLPO into a single stage transition $\Sigma_t \rightarrow \Sigma_{t+1}$ while effectively preserving the computational complexity. 
% Mathematically we impose this in the MDP formulation by appropriately restricting the set of all actions allowed at every state. 
% Fur, we now have flexibility of associating distinct unknown parameters with the same facility. 
% Infinite horizon problems represent a reasonable approximation of problems involving large state and action spaces, with comparatively straighforward implementation of stationary optimal policies \cite{bertsekas1996neuro}. One of our previous studies \cite{srivastava2021parameterized} demonstrated a framework for a parameterized infinite horizon Markov Decision Process (MDP), where states, actions, and costs are influenced by specific parameters. Consequently, we undertake a topographical transformation of the original FLPO problem to reframe the finite-horizon MDP formulation within an infinite horizon framework.
 
\subsection{Infinite-Horizon Parametrized MDP Model of FLPO}

The augmented FLPO now can be modeled as an infinite horizon discounted parametrized SDM that consists of a cost free terminal state $\delta$, the state-space
% The MDP is formally defined as a tuple $\left\langle\mathcal{S}, \mathcal{A}, c, p, \gamma \right\rangle$, where 
    % \begin{align*}
    %     \mathcal{S} = \{& n_1, n_2, \dots, n_N, \\
    %     & f_1^1, f_2^1, \dots, f_M^1, \\ 
    %     & f_1^2, f_2^2, \dots, f_M^2, \dots, \\ & f_1^M, f_2^M, \dots, f_M^M, \delta\}
    % \end{align*}
$       
        \Sinf = \Gamma_0 \cup \Gamma_1 \cup \dots \cup \Gamma_{M+1}
$
% Note that this enable us to determine the optimal value of stage-dependent parameters in the original FLPO in a single-stage transition of the stacked FLPO framework. 
and the action-space
    % \begin{align*}
    % \mathcal{A} = \{& f_1^1, f_2^1, \dots, f_M^1, \\ 
    % & f_1^2, f_2^2, \dots, f_M^2, \dots, \\ & f_1^M, f_2^M, \dots, f_M^M, \delta\}
    % \end{align*}
    $
        \Ainf = \Gamma_1 \cup \Gamma_2 \cup \dots \Gamma_{M+1}
    $.
An element $a_t \in \Ainf, a_t = f^k_j$ or $a_t = \delta$ indicates the direction of the agent at the current state $s_t \in \Sinf$ to the facility or destination represented by $a_t$. 
Denote $\curlyb{\xi_s}$ and $\curlyb{\lambda_a}$ as state and action parameters respectively, which can be known or unknown and we may have to optimize the cost accordingly. 
For the simplest FLPO problem, there are no action parameters, and $\curlyb{\xi_s}$ includes fixed node locations, destination locations and unknown facility locations. 

Define $\forall k = 0,1,\dots,M+1, \Gamma_k(\Sinf) \subset \Sinf$ and $\Gamma_k(\Ainf) \subset \Ainf$ as a subset of state and action spaces respectively corresponding to $\Gamma_k$.
For $s\in \Sinf$, denote $\Ainf(s)$ as the set of all possible actions available at state $s$, then to ensure the FLPO stage-wise transition constraints, we have $\forall s \in \Gamma_0(\Sinf), \Ainf(s) = \Gamma_1(\Ainf)$ and $\forall k =1,2,\dots,M, \forall s \in \Gamma_k(\Sinf), \Ainf(s) = \Gamma_{k+1}(\Ainf)\cup \curlyb{\delta}$.
% Each action $a \in \mc A $ has an associated parameter $\eta_a$ for which we may have to optimize the network. 
% \textit{Similar to the state parameters, this enables us to determine the optimal value of stage-dependent action parameters in a single-stage transition}.
% Further, $c:\mathcal{S}\times\mathcal{A} \times \mathcal{S} \rightarrow \mathbb{R}$ is the cost of transition from the state-action pair $(s_t, a_t)$ to the state $s_{t+1}$, $p:\mathcal{S} \times \mathcal{S} \times \mathcal{A} \rightarrow \left[0,1\right]$ represents the probability of transition from a state-action pair $(s_t, a_t) \in \mathcal{S}\times\mathcal{A} $ to a state $s_t\in\mathcal{S}$, $\gamma \in (0,1]$ is the discounting factor. 
A control policy $\mu_i:\Sinf\times\Ainf \rightarrow \left\{0,1\right\}$ determines the action $a \in \Ainf$ taken at state $s\in \Sinf$ if $\mu_i(a|s) = 1$ otherwise $\mu_i(a|s) = 0$. 
The state-transition probability $p_i:\Sinf \times \Ainf \times \Sinf \rightarrow \rectb{0,1}$ and the policy $\mu_i$ are such that $\forall s\in \Gamma_k(\Sinf), p_i(s'|s,a) = 0, \forall a \in \Ainf(s), \forall s' \in \Sinf / \roundb{\Gamma_{k+1}(\Sinf)\cup \curlyb{\delta}}$ and $\mu_i(a|s) = 0,  \forall a \in \Ainf / \Ainf(s)$
    % \begin{enumerate}
    %     \item $\forall s\in \Gamma_k(\mc S)$, $\forall a \in \mc A(s)$, the agent can only transition to $s' \in \Gamma_{k+1}(S) \cup \curlyb{\delta}$
    %     \begin{align*}
    %         \sum_{s'\in \Gamma_{k+1}(\mc S)\cup\{\delta\}}p(s'|s,a) = 1 
    %     \end{align*} \label{cons:policy_and_model1}
    %     \item Further, $\forall s \in \Gamma_k \subset \mc S$, the support of the policy $\mu(\cdot|s)$ is the set $\mc A(s)$ as defined earlier
    %         \begin{align*}
    %             \sum_{a \in \mc A(s)}\mu(a|s) = 1, \forall k = 0,1,\dots,M
    %         \end{align*}
    %         \label{cons:policy_and_model2}
    %     % \item If $s = \delta$, then $\mu(a|s) = 1$, if $a = \delta$ and $\mu(a|s) = 0$ otherwise.
    %     \label{cons:policy_and_model3}
    % \end{enumerate}

The resulting optimization problem is to find the optimal policy $\mu_i^*$ and the parameters $\curlyb{\zeta_s}$ such that the state value function is minimized $\forall s \in \Gamma_0(S)$ 
\begin{align*}
    J^{\mu_i}_{\xi\lambda}(s) = \mb E_{p_{\mu_i}}\rectb{\sum_{t=0}^\infty \gamma^t c_i\roundb{x_t(\xi), u_t(\lambda), x_{t+1}(\xi)}|x_0 = s}, 
\end{align*}
where the expectation is taken over the distribution $p_{\mu_i}(\cdot|x_0)$ over the set of all the paths generated by the MDP.
% \begin{align*}
%     \omega_i = \{u_0, x_1, u_1, x_2, u_2, \dots\}, u_t, x_t \in \Ainf, \Sinf, \forall t \in \mb Z_{\geq 0}
% \end{align*}
This is identical to a paraSDM problem (as described in \ref{subsec:parasdm_timeinvariant}) and is solved using the DA approach resulting into the following policy with additional constraints on the states and action spaces, we have, $\forall s\in \Sinf, a \in \Ainf(s)$
% Randomizing the policies $\mu_i$, and then applying the Maximum Entropy Principle to find the policies results into a solution of the form (\ref{eq:policy_gibbs}) and (\ref{eq:stateaction_contraction}). Hence, $\forall a \in \Ainf(s), \forall s \in \Sinf$ we have
\begin{align}
    \mu^*_{\beta}(a|s) & = \frac{e^{-\frac{\beta}{\gamma}\Lambda_\beta\roundb{s,a}}}{\sum_{a'\in \Ainf(s)}e^{-\frac{\beta}{\gamma}\Lambda_\beta\roundb{s,a'}}}, \nonumber \\
    \Lambda_\beta(s,a) & = \sum_{s'} p_{ss'}^a \rectb{\bar{c}_{ss'}^a -\frac{\gamma^2}{\beta} \log \sum_{a' \in \mc A_i(s')} e^{-\frac{\beta}{\gamma}\Lambda_\beta(s',a')}},  \nonumber
\end{align}
and the gradients are obtained using the fixed point iterations of the form  (\ref{eq:gi_stateparams}). However, instead of implementing the gradient descent scheme, we use the BFGS solver that results in faster convergence to the solution to the optimal value of parameters. 
% Further, this enables us to incorporate 

\begin{figure*}[tbhp!]
    \centering
    \includegraphics[trim={1cm 1cm 1cm 1cm},clip,width=2\columnwidth]{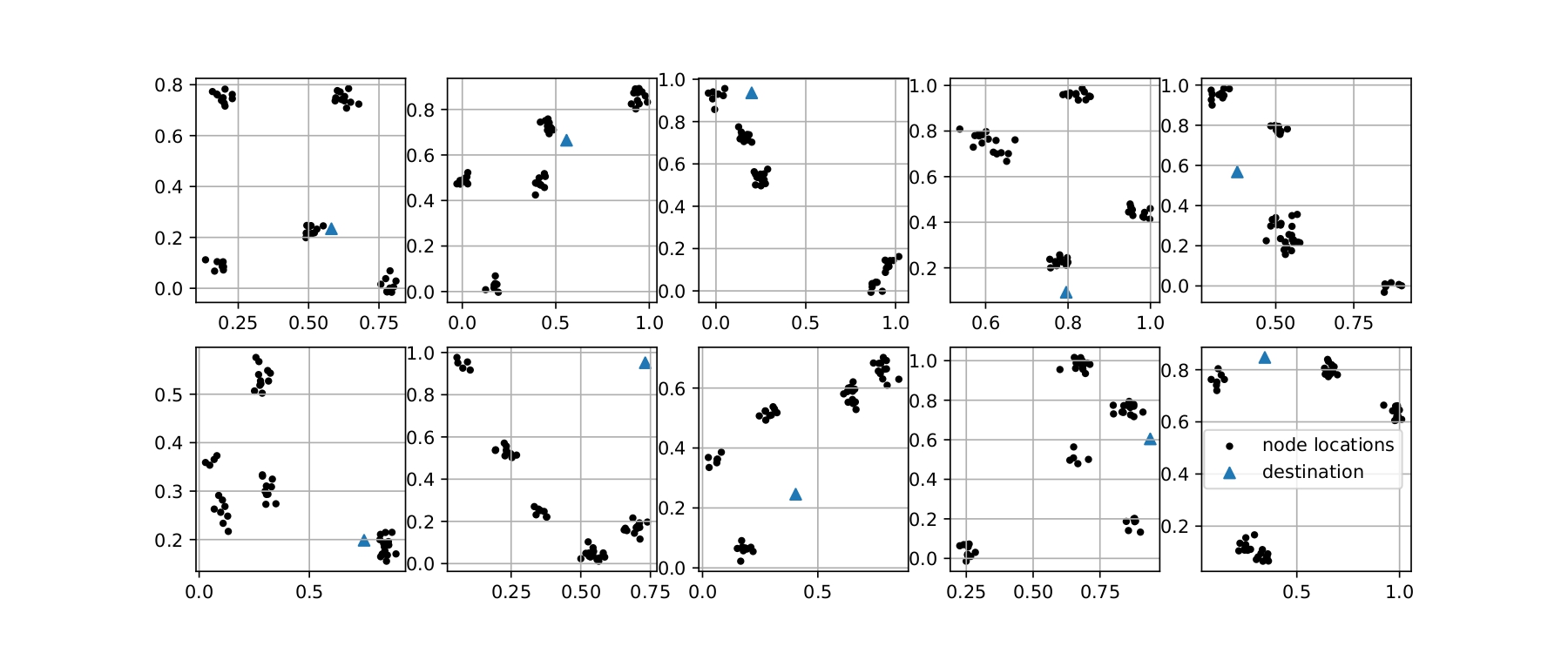}
    \caption{$x-y$ coordinates of the datasets generated for 5G small-cell simulations shown in 2D plane. Each figure depicts distributed user nodes as black points and the destination center as a solid blue triangle. Additionally, all coordinates are normalized to fit within a unit square, facilitating efficient hyperparameter tuning.}
    \label{fig:datasets}
\end{figure*}

% \subsection{Towards enabling learning}
One of the key advantages of viewing FLPO as an infinite-horizon SDM is the ability to incorporate machine learning techniques. This approach facilitates scaling the original FLPO framework to accommodate large state and action spaces, which is especially beneficial in environments where the state-transition probability $p_{ss'}^a$, the cost function $c_{ss'}^a$ and its dependence on the parameters $\curlyb{\xi}$ and $\curlyb{\lambda}$ is unknown. 
This can be achieved by either learning the state-action value function and value function gradients with respect to parameters based on Q-learning, deep-Q learning, or by learning the optimal policies and optimal parameters based on policy iteration methods.
Q-learning is useful to approximate $\Lambda_{\beta,\xi\lambda}^{\mu}(s,a)$ and the gradients $G_{\xi_s}^\beta(s')$ and $G_{\lambda_s}^\beta(s')$ at every $\beta$ using the following stochastic iterative updates
\begin{align}
    & \Psi_{t+1}(x_t,u_t) = (1-\nu_t(x_t,u_t))\Psi_t(x_t,u_t) \nonumber \\ \nonumber & \quad + \nu_t(x_t,u_t)\rectb{c_{x_tx_{t+1}}^{u_t} - \frac{\gamma^2}{\beta}\log \sum_{a'\in\mc A(x_t)}e^{-\frac{\beta}{\gamma}\Psi_{t}(x_{t+1},u_t)}}, \\ \nonumber
    & K^{t+1}_{\alpha}(x_{t},u_t) = 
    \roundb{1-\nu_t(x_t,u_t)} K^t_{\alpha}(x_{t},u_t) \\ \nonumber
    & \quad \quad \quad \quad \quad
    + \nu_t(x_t,u_t) \rectb{\frac{\partial c_{x_tx_{t+1}}^{u_t}}{\partial\alpha} + \gamma G^t_\alpha(x_{t+1})},
\end{align}
where $\nu_t(x_t,u_t) \in \left(0,1\right]$ is a step size parameter and $G^t_\alpha(x_{t+1}) = \sum_{a\in\mc A(x_{t+1})} \mu_{a|x_{t+1}} K^{t+1}_{\alpha}(x_{t},a), \alpha \in \curlyb{\xi_s, \lambda_a}$. The function $\Psi_{t+1}(x_t,u_t)$ converges to the fixed point of optimal $\Lambda_{\beta,\xi\lambda}^*$ under specific conditions on $\nu_t$ \cite{srivastava2021parameterized}. Further, the parameter values can then be updated using gradient descent schemes or standard numerical optimization methods. However, tabular Q-learning is infeasible for scenarios with large states and actions. Deep Q-learning is devised to approximate the state-value and gradients using Deep Neural Networks (DNNs) for large or continuous state spaces. For example, a neural network  $\hat{\Lambda}_{\beta,\xi\lambda}(s,a; w)$ for state-action value function can be learned by minimizing the following loss by iteratively interacting with the environment
\begin{align*}
    L_i(w_i) = \mb E_{s,a,c,s'\sim \tau(\cdot)} \rectb{\roundb{y_i - \hat{\Lambda}_{\beta,\xi\lambda}(s,a; w_i)}},
\end{align*}
where $y_i = c(s,a,s') - \frac{\gamma^2}{\beta}\log \sum_a e^{-\frac{\beta}{\gamma}\hat{\Lambda}_{\beta,\xi\lambda}(s',a'; w_{i-1})}$, and $\rho$ is a distribution over the transitions $s,a,s',c$ observed in the environment \cite{srivastava2021parameterizedthesis}.

One of our ongoing works is showing an on-policy, policy optimization algorithm under the Maximum Entropy Principle that guarantees monotonic policy improvement with iterations for model-free reinforcement learning for large state as well as large action spaces. It also aims to provide a sample-based algorithm to approximate policies using neural networks along with monotonic convergence guarantees. By incorporating these advancements into our current framework, we will be equipped to effectively address a wide range of problems in parametrized sequential decision making.

\begin{figure}[tbhp!]
    \centering
    \includegraphics[width=0.9\columnwidth]{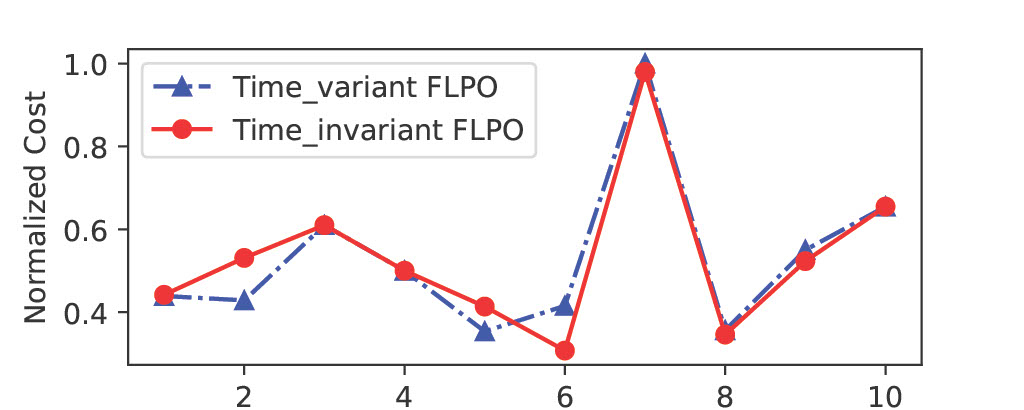} \\
    \includegraphics[width=0.9\columnwidth]{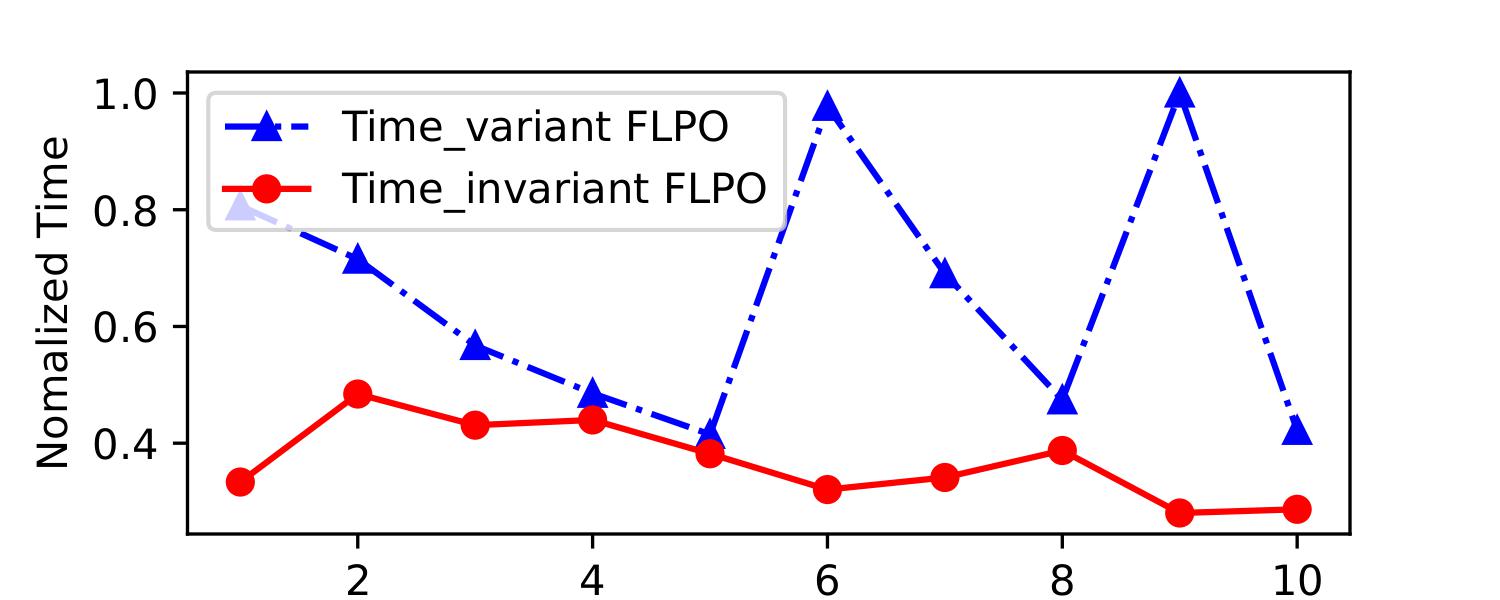} \caption{The normalized cost and performance time comparison of the base time-invariant FLPO solution and the proposed time-invariant FLPO solution.}
    \label{fig:results}
\end{figure}

{\color{black}
\section{SIMULATIONS AND RESULTS}

To verify whether the solution of the proposed non-stationary MDP is equivalent to the solution of the base FLPO framework of \cite{srivastava2020simultaneous}, a number of carefully designed simulations were conducted , and based on final values of optimal decision variables and cost function, the solutions were compared. 

As for the simulation setup, we choose the small cell network problem in which the data packets from spatially distributed user nodes should be transported to a destination center through a network of cell towers. We determine the cost of sending the data packets between the nodes, cell-towers and destination centers as the corresponding squared-Euclidean distance between them. The objective is to find the optimal location of the cell-towers and an optimal route from each user node to the destination center such that the cumulative cost over all the user nodes is minimized. This is a quite difficult problem to solve, as the number of decision variables for the policy is combinatorially large $\mathcal{O}(\sum_{k=1}^M \bigl(\begin{smallmatrix} M \\ k \end{smallmatrix}\bigr) k!)$, and on top of that $M$ facility locations are need be be optimized simultaneously. In fact, we have shown in \cite{ourAIAA} that standard optimization algorithms, such as the Genetic Algorithm (GA), Particle Swarm Optimization (PSO) and enhanced Covariance Matrix Adaptation Evolution Strategy (CMA-ES) which are commonly used in the literature fail to address these kind of problems efficiently, even on networks with a few number of facilities and $N \approx 10$. Furthermore, our future objective is to enhance the scalability of our solutions to accommodate problem sizes the number of facilities $N > 100$, thus emphasizing the importance of the learning aspect discussed earlier.
% used for the distance metric as a proper candidate to verify the methodology of this paper. 

With that said, In ten different scenarios (Figure \ref{fig:datasets}), a total of $N=50$ nodes are randomly scattered with a Gaussian distribution in 5 different unequally sized clusters, and for each scenario a set of $M=5$ facilities are required to be placed optimally to route the nodes to the destination point. In different scenarios, the means of node clusters are placed at various relative positions with regards to the destination point, and we have chosen $C_j = 0.0005\mathbf{I}_{2 \times 2}, \ j \in \{1,\dots M\}$ as the covariance of each cluster, where $\mathbf{I}_{2 \times 2}$ denotes the identity matrix. In some cases, the nodes should optimally decide to go to the destination point directly, while others need to take routes along facilities first.
Figure \ref{fig:results} shows the results of the simulation where the normalized optimal cost value and performance times are compared for the time-variant FLPO and the proposed time-invariant FLPO solution. The cost plot demonstrates that the proposed time-invariant solution achieves near-identical performance to the original FLPO solutions, and in some instances, even surpasses it. In a few cases however, the original FLPO solution provides a slightly lower cost by providing a different solution from the proposed method; one such case (dataset number 2) is shown in Figure \ref{fig:solution comparison}.
\begin{figure}[tbhp!]
    \centering
    \includegraphics[trim={0.8cm 1.3cm 1.2cm 1cm},clip,width=0.65\columnwidth]{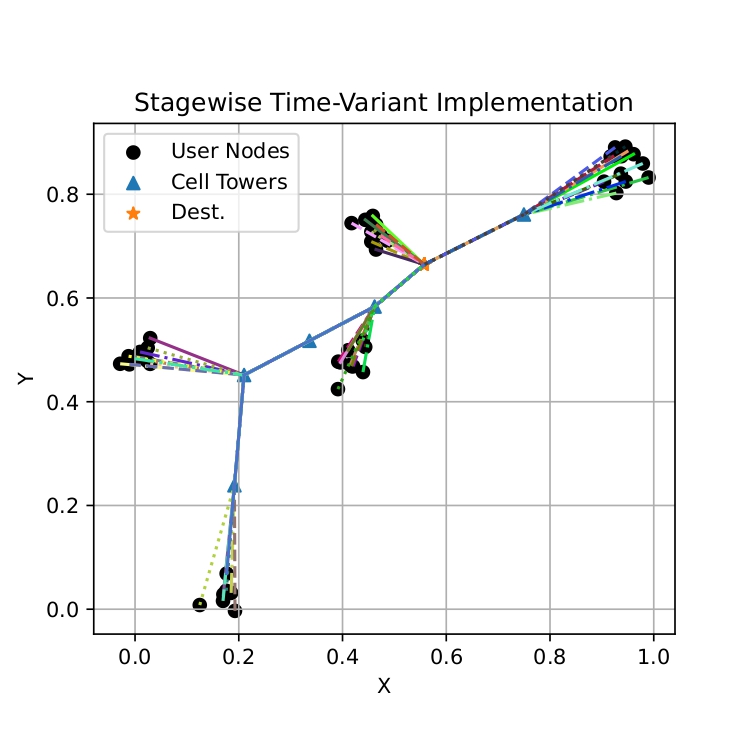} \\ \includegraphics[trim={0.8cm 1.3cm 1.2cm 1cm},clip,width=0.65\columnwidth]{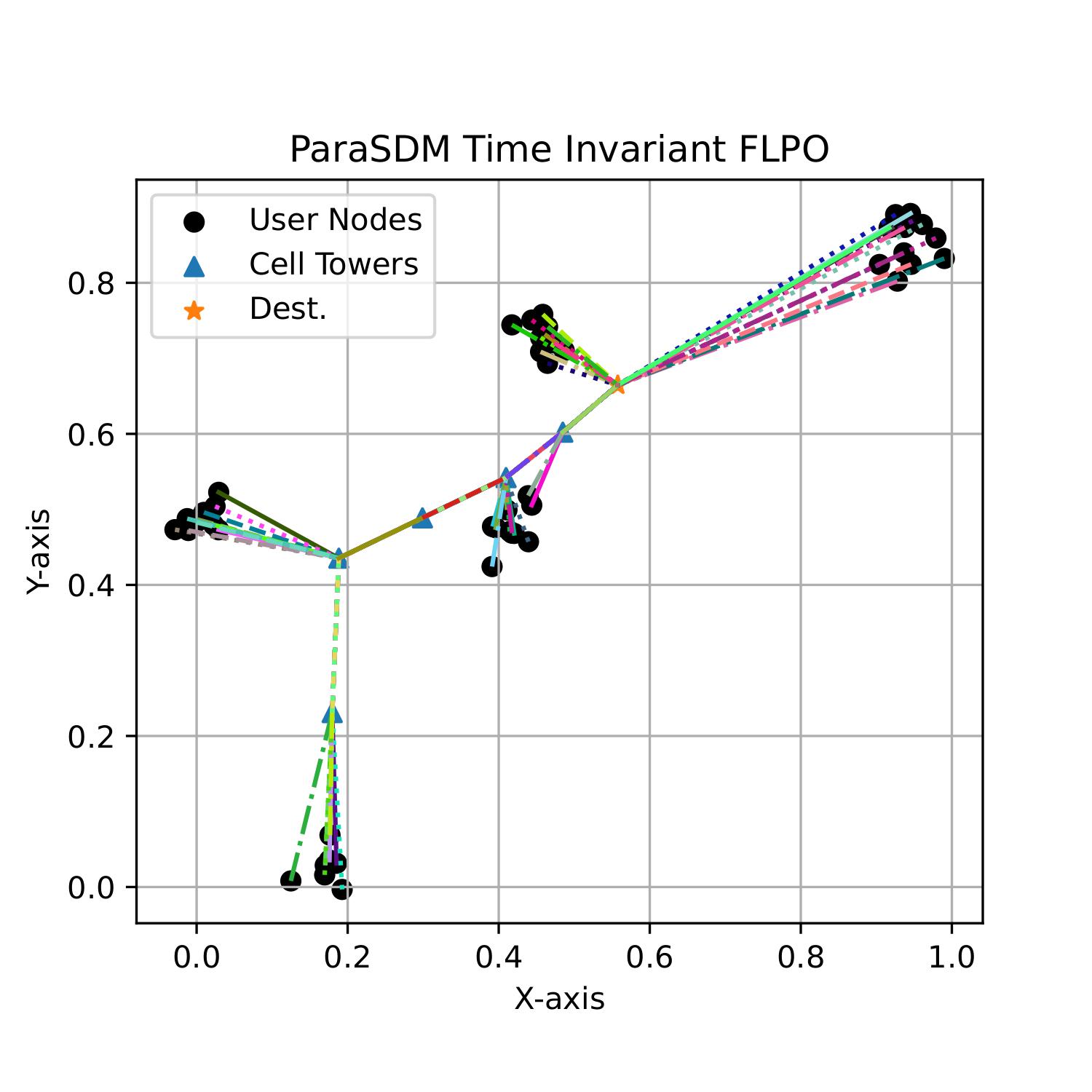}
    \caption{Comparison of the solutions for the dataset 2, where optimal location of the facilities (Cell Towers) and the routes of user nodes are shown.}
    \label{fig:solution comparison}
\end{figure}

The time plot on the other hand, reveals a significant advantage in using the time-invariant methodology, as anticipated in the introduction. This advantage stems from the ability to directly compute the Jacobians of the value function with respect to state parameters. 
% ({\color{red}{I believe this claim becomes dicey now since the paraSDM one takes longer time for a larger problem?}}{\color{blue} I'm not sure, this could be due to implementation issues. maybe we can make a softer claim?}{\color{red}{How about we add that we will investigate large size problems?}})
In contrast, the original FLPO solution requires more cumbersome calculations for obtaining derivatives. This facilitates direct optimization of the value function, leading to faster and more efficient learning strategies, particularly for large network sizes where the time-invariant nature of the proposed methodology becomes even more beneficial. %{\color{red} any other reason why the cost/time of the para-sdm FLPO is better?} {\color{blue}{I think this is it. Another plus point we can mention - in the cases where parameters are unknown, and costs are also unknown, the G-Iteration can be replaced with QLearning with DNNs in it. Basically, we will be able to learn the gradient too. This is more like future work but can be mentioned citing Amber's work?} \color{green} isn't this mentioned in the introduction?
}
% {\color{red}{Should we also attempt to show simulations by varying the discounting factor?} \color{blue} I don't think that would be useful}
% \subsection{explain the simulation setup, and why it was chosen - Small Cell Network Problem, UAV Problem AIAA}
% \subsection{explain the parameters chosen for that setup, for example number of nodes and facilities, and the reason behind choosing those values}
% \subsection{explain what metrics were used to compare the solutions - i) Cost value, ii) Difference in the final location of all the facilities, iii) routes can be demonstrated visually by numbering all the nodes and facilities in the graph, iv) time taken for the solution}}

\section{CONCLUSION AND ONGOING WORK}
% FLPO problems can be treated as time-varying finite horizon sequential decision making problems. In this paper, we have reinterpreted a time-variant finite-horizon FLPO as a time-invariant paraSDM problem by transforming a multi-stage architecture of the FLPO into a single-stage. Our results show the equivalence between the two viewpoints and the infinite horizon paraSDM performs better in terms of the computation time and also the final costs. Further, the infinite horizon perspective of FLPO provides us with additional flexibilities - such as incorporating time varying parameters into a single stage, expanding the scope to the problems where the transition probabilities are not deterministic and further make the problem scalable as now we can use neural network architecture to learn the optimal policies and the value function.

Our study highlights that time-varying FLPO problems can be approached as time-invariant decision-making tasks. By reinterpreting a time-variant finite-horizon FLPO as a stationary policy ParaSDM problem, we simplify the complexity into a single step along with the topological constraints that preserve the allowed transitions in the stage-wise framework of FLPO. Our findings indicate that while both perspectives are essentially equivalent, the infinite horizon ParaSDM approach for FLPO demonstrates better performance in computational efficiency and final cost results. This perspective also offers additional advantages, such as the ability to integrate time-varying parameters into a single stage and expand applicability to scenarios with uncertainty in models. 
Furthermore, it enables scalability by leveraging learning approaches to learn optimal strategies and value functions.


%%%%%%%%%%%%%%%%%%%%%%%%%%%%%%%%%%%%%%%%%%%%%%%%%%%%%%%%%%%%%%%%%%%%%%%%%%%%%%%%

%%%%%%%%%%%%%%%%%%%%%%%%%%%%%%%%%%%%%%%%%%%%%%%%%%%%%%%%%%%%%%%%%%%%%%%%%%%%%%%%

%%%%%%%%%%%%%%%%%%%%%%%%%%%%%%%%%%%%%%%%%%%%%%%%%%%%%%%%%%%%%%%%%%%%%%%%%%%%%%%%

%%%%%%%%%%%%%%%%%%%%%%%%%%%%%%%%%%%%%%%%%%%%%%%%%%%%%%%%%%%%%%%%%%%%%%%%%%%%%%%%

\bibliographystyle{IEEEtran}
\bibliography{mybibfile}

\end{document}